\documentclass[11pt, letterpaper, times, onecolumn, preprint]{aastex631}
\usepackage{nopageno}

\usepackage[sort&compress]{natbib}
\bibpunct{(}{)}{;}{}{,}{,}

\usepackage{graphicx} 
\usepackage{epstopdf} 
\usepackage{wrapfig}
\usepackage{framed}
\usepackage{enumitem}
\setlist[itemize,0]{leftmargin=*,itemsep=0pt}
\setlist[enumerate,0]{leftmargin=*,itemsep=0pt}
\setlist[description]{leftmargin=*,itemsep=0pt}

\usepackage{amsmath}
\usepackage{amssymb}
\usepackage{epsfig}
\usepackage{graphicx}
\usepackage{tabularx}
\usepackage{boldline}
\usepackage{slashed}
\usepackage{diagbox}
\usepackage{color}
\usepackage[normal]{subfigure}
\usepackage[utf8]{inputenc}
\usepackage{colortbl}
\usepackage{wrapfig}
\usepackage{verbatim}
\usepackage{verbatim}

\usepackage{booktabs, makecell, multirow, xspace}
\newcolumntype{L}[1]{>{\raggedright\let\newline\\\arraybackslash\hspace{0pt}}p{#1}}
\newcolumntype{C}[1]{>{\centering\let\newline\\\arraybackslash\hspace{0pt}}p{#1}}
\newcolumntype{R}[1]{>{\raggedleft\let\newline\\\arraybackslash\hspace{0pt}}p{#1}}

\usepackage{url}
\usepackage[normalem]{ulem}
\usepackage{amsmath}
\usepackage{comment}
\usepackage{svg}
\usepackage{natbib}
\usepackage{fontawesome}

\usepackage[most]{tcolorbox}
    \tcbuselibrary{theorems}
    \newtcolorbox{answerbox}{sharp corners=all, colframe=black, colback=black!5!white, boxrule=1.5pt, width = 1\textwidth, valign=center}

\definecolor{md_orange}{rgb}{1,0.55,0.1}
\definecolor{md_blue}{rgb}{0.1,0.55,1}
\definecolor{md_red}{rgb}{1,0.1,0.1}

\newtcolorbox{overview}[2][]{
    colback=samcolor!5!white,
    colframe=samcolor!75!black,
    colbacktitle=samcolor!10!white,
    coltitle=samcolor!70!black,
    title={#2},fonttitle=\bfseries,#1}

\newtcolorbox{orangebox}[2][]{
    colback=md_orange!5!white,
    colframe=md_orange!75!black,
    colbacktitle=md_orange!10!white,
    coltitle=md_orange!70!black,
    title={#2},fonttitle=\bfseries,#1}

\newtcolorbox{bluebox}[2][]{
    colback=md_blue!5!white,
    colframe=md_blue!75!black,
    colbacktitle=md_blue!10!white,
    coltitle=md_blue!70!black,
    title={#2},fonttitle=\bfseries,#1}
    
\newtcolorbox{redbox}[2][]{
    colback=md_red!5!white,
    colframe=md_red!75!black,
    colbacktitle=md_red!10!white,
    coltitle=md_red!70!black,
    title={#2},fonttitle=\bfseries,#1}

\newtcolorbox{graybox}[1][]{
    colback=green!5!white,
    colframe=green!75!black,
    colbacktitle=green!10!white,
    coltitle=green!70!black,
    #1}

\usepackage{makecell}
\usepackage{todonotes}
\definecolor{gold}{RGB}{255,215,0}

\newcommand{\Penn}{\affiliation{Department of Physics \& Astronomy, University of Pennsylvania, Philadelphia, PA 19104, USA}}
\newcommand{\Dartmouth}{\affiliation{Department of Physics \& Astronomy, Dartmouth College, 6127 Wilder Laboratory, Hanover,
NH, USA}}
\newcommand{\Berkeley}{\affiliation{Astronomy Department, University of California, Berkeley}}

\newcommand{\OSU}{
\affiliation{Center for Cosmology and Astroparticle Physics, The Ohio State University, 191 West Woodruff Avenue, Columbus, OH 43210, USA}
\affiliation{Department of Physics, The Ohio State University, 191 West Woodruff Avenue, Columbus, OH 43210, USA}
\affiliation{Department of Astronomy, The Ohio State University, 140 West 18th Avenue, Columbus, OH 43210, USA}}
\newcommand{\CfA}{\affiliation{Center for Astrophysics | Harvard \& Smithsonian, Cambridge, MA 02138, USA}} %define a command for your institution in this doc

\shorttitle{Roman Early Definition Science}
\shortauthors{Sanderson et al.}

\begin{document}

\correspondingauthor{Robyn E. Sanderson}
\email{robynes@sas.upenn.edu}

\title{Recommendations for Early Definition Science with the Nancy Grace Roman Space Telescope}

\author[0000-0003-3939-3297]{Robyn E. Sanderson}
\Penn
\author[0000-0003-1468-9526]{Ryan Hickox}
\Dartmouth
\author[0000-0002-2951-4932]{Christopher M. Hirata}
\OSU
\author[0000-0002-1139-4880]{Matthew J. Holman}
\CfA
\author[0000-0001-9611-0009]{Jessica R. Lu}
\Berkeley
\author[0000-0002-5814-4061]{V. Ashley Villar}
\CfA
%add yourself here; put affiliations in affiliations.tex

\begin{abstract}
The Nancy Grace Roman Space Telescope (Roman), NASA's next flagship observatory, has significant mission time to be spent on surveys for general astrophysics in addition to its three core community surveys. We considered what types of observations outside the core surveys would most benefit from early definition, given 700 hours of mission time in the first two years of Roman's operation. We recommend that a survey of the Galactic plane be defined early, based on the broad range of stakeholders for such a survey, the added scientific value of a first pass to obtain a baseline for proper motions complementary to Gaia's, and the significant potential synergies with ground-based surveys, notably the Legacy Survey of Space and Time (LSST) on Rubin.
We also found strong motivation to follow a community definition process for ultra-deep observations with Roman. 
%Roman will be an amazing telescope; we considered what sort of observations would most benefit from early definition; here are our findings.
\end{abstract}

\section{Context} %RES
\label{sec:context}
The Nancy Grace Roman Space Telescope (Roman) will spend at least one quarter of its prime
mission time (1.25 out of the first 5 years) on Astrophysics Surveys that have yet to be
defined. These Astrophysics Surveys are in addition to the three Core Community Surveys (CCSs)
targeted at answering specific questions on dark energy and exoplanet demographics. 
%but willproduce datasets that also enable investigations on a broad array of other subjects.

Some potential Astrophysics Surveys may benefit significantly from being selected early, to enable multi-year preparatory activities that can enhance the value of the survey and its data products. To investigate whether such an Early-Definition Astrophysics Survey (EDAS) would be beneficial, the Roman Project released a ``Roman Early-Definition Astrophysics Survey Opportunity:
Request for Information'' (RfI) to solicit comments from the science community on (a) whether to pre-select any Astrophysics Survey, and (b) to outline and submit survey concepts that would demonstrably benefit from selection as an EDAS. The request instructed the community to submit a high-level conceptual description of a proposed survey  using the Wide Field Instrument, to be executed within the first two years of the Roman mission, using up to $\sim700$ hours of wall clock time (the equivalent of one
full month). The RfI received twenty responses with over 340 unique authors, demonstrating the community's enthusiasm for the possibilities offered by early definition.

The Roman Project charged a committee made up of the authors of this document to review the results of the RfI and (1) recommend whether to pre-select one or more proposed surveys for early definition, and (2) if pre-selection is recommended, provide a ranked list of potential survey themes. Full survey specifics and detailed observational strategies would be defined through a future community process similar to that now underway for the CCSs. A recommended Roman EDAS would thus be defined in parallel with the CCSs, so that all three plans will be known in advance of Cycle 1.

\begin{bluebox}{Charge to the Roman Early Definition Assessment Committee}
    The Committee should review all the submissions and for any that include a survey concept, assess the degree to which the submission addresses the above threshold and scientific merit criteria. Following their review of submission, the Committee shall provide an assessment as to whether to proceed with specifying a Roman Early-Definition Astrophysics Survey, and if so, the Committee shall provide a ranked list of survey themes. The Committee may provide additional narrative findings as necessary on the topic of executing one or more Early-Definition Astrophysics Surveys. The Committee also may choose to request additional science or technical details on one or more of the proposed survey concepts.

The Committee shall write a brief report, to be made publicly available, providing the motivation
for their findings. The NASA Roman Senior Project Scientist will make the final decision to
accept, reject, or amend the findings, and may request input from additional groups or
individuals before doing so.
\end{bluebox}

\section{Rationale for Early Definition}
\label{sec:rationale}

The committee identified several universal arguments for early definition of a Roman survey. Foremost among these, perhaps, is the potential for engaging and mobilizing a community of scientists to prepare to analyze Roman data. The process for defining a putative Roman EDAS will involve all of the community, distinct from the individually competed proposals in standard observing cycles. There is no anticipated restriction on the size or breadth of teams that can propose Roman Astrophysics Surveys in the standard observing cycles, so selecting an EDAS is not necessarily the \emph{last} chance for community-driven definition of a Roman survey, but is a \emph{guaranteed} opportunity to do so (see 2(d) in the boxed RfI criteria). Since a breadth of community input ensures a breadth of scientific impact (especially for the potential surveys that we ranked highly), the committee considered this potential for community-driven definition to be the most important ``opportunity'' (in the sense of criterion 1 in the RfI) that would be lost by not defining a survey early.

Another particular advantage of early mobilization for many science cases is the ability to plan for and acquire data from other telescopes that makes certain Roman analyses possible or greatly improved. In particular, arranging for complementary observations from telescopes with high oversubscription factors, such as spectrographs on 8- and 10-meter-class telescopes, or for particular timing, such as simultaneous or near-simultaneous observations with multiple facilities, was considered a compelling reason for early definition. One example is the Subaru Prime Focus Spectrograph, on which the Roman project does have guaranteed time but not until the latter half of its five-year mission. Likewise, the ability to influence survey strategy for other facilities by defining a survey with Roman was considered compelling.

Roman's total amount of observing time is of course finite, so selecting an EDAS will necessarily take observing time away from other programs. The committee assumed that this time would come from the proportion made available for general Astrophysics Surveys, but also recognized that allocations between CCSs and GA are yet to be determined. A Roman EDAS would not be a \emph{core} community survey, but would be community defined and have time allocated alongside the other Roman programs.

The committee considered the possibility of engaging with the public through images that showcase Roman's unique capabilities---its wide field of view, exquisite resolution, and combination of infrared bandpasses---as an argument for early definition. An enormous amount of compelling science is, of course, already planned in the Roman Core Community Surveys, but much of this core science will take some time to come to fruition. There will likely be a significant mismatch between the timescale for obtaining spectacular large-format images, which will be very fast (hours) with Roman's wide field of view and agility, and that for the substantial science output from the types of surveys described in the submitted whitepapers, which represent the best use of Roman as a scientific instrument. The first public images will likely be completed as part of commissioning and tests early in the mission. The committee thus regarded the potential for public engagement as nice to have, but not sufficient motivation in itself for early definition.

In many whitepapers the need for early data was cited as an argument for early definition. In most cases, as outlined below, this argument really pertained to early \textit{execution}, i.e. making observations close to the beginning of Roman's operation. This is distinct from early \textit{definition}, in which the fields are known in advance but observations could be executed at any time during the mission. Thus the committee considered early execution alone to be insufficient justification for early definition (the competed Cycle 1 will provide an opportunity to argue for early execution) but rather as an important supporting argument. Cases where defining the survey would best be done via a community process, that \emph{also} require early execution, were considered to have a particularly compelling justification for early definition.

Since the RfI, the funding stream for Roman Wide-Field Science investigations has also been established. This program, which included 18 funded projects in 2022 and will have another call for additional proposals in 2024, was seen by the committee as a way to support development of software, simulations, and theory for potential applications to Roman observations. Notably, this call does \emph{not} require that the target observations have been defined. Thus, this type of work \emph{on its own} was not seen as a compelling primary argument for early definition, although many science cases put forward in whitepapers that the committee considered compelling for the other reasons discussed above included this as a plausible secondary rationale.

The committee additionally considered the criteria in the box below, which were provided to the community during the request for whitepapers.
\begin{bluebox}{Criteria provided in the Request for Information}
\begin{enumerate}
    \item  The science enabled by the proposed survey must be significantly
enhanced by specific preparatory activities that are enabled by early selection and definition of the
proposed survey concept. These may include, e.g., supporting facility observations, software
development work, theoretical/simulation efforts, etc. This must present an opportunity that would
otherwise be lost by waiting until the first Call for Proposals and subsequent competitive peer review.

\item Additionally, the proposed survey concept:
\vspace{-6pt}
\begin{enumerate}
    \item must enable investigations yielding significant scientific advancements,
commensurate with the allocation of up to one month of observing time on a major NASA Observatory;
    \item  must exploit the unique observational capabilities of Roman and must be beyond the capabilities of the ground-based and space-based datasets expected to be available at the time of Roman launch;
    \item must be distinct in design from the Core Community Surveys that Roman will conduct and must enable science investigations that will not already be possible with the Core Community Survey data;
    \item should create datasets that will allow a broad segment of the astronomical community to pursue a wide range of investigations across range of subject areas.
\end{enumerate}
\end{enumerate}
\end{bluebox}

\section{Main survey themes}
\label{sec:themes}
The responses to the RfI, which are listed in Table \ref{tab:whitepapers}, fell into a handful of general themes:
\begin{description}
    \item[Deep fields of galaxies] Proposals to observe a total sky area of 0.3--1 sq. deg at a depth of 30${}^{\mathrm{th}}$ magnitude or greater. The science goals of these proposals broadly covered galaxy formation at high redshift, the Epoch of Reionization, and active galactic nuclei at intermediate redshift.  
    \item[Deep fields with grism] As above, but with additional observations using the grism for deep multiplexed spectroscopy.
    \item[Intermediate ``wide'' fields] Proposals to observe 20--50 sq. deg on the sky at a depth of $\sim$27${}^{\mathrm{th}}$ magnitude, often with the aim of significant synergy with other surveys or observations. A significant subset of submissions proposed using Roman's K band to obtain redder images than Euclid or Rubin will collect, in order to study galaxy formation with large samples at high redshift, or to otherwise augment surveys in other wavelength ranges.
    \item[High time-cadence] Repeat observations of one or a few Roman fields of view at a cadence of one minute or less, much faster than planned for any of the the Core Community Surveys. These proposals focused on science related to star formation, stellar evolution, and planet formation in a variety of environments. 
    \item[Galactic plane] Several submissions argued for surveying the Galactic plane as a significant discovery space with Roman, due to the unprecedented combination of resolution, depth, and wavelength range for such observations. The proposed surveys would use the F106 and/or F129 and F213 Roman bands to depths of 23--25.5 mag over about 1000 square degrees, and advocated for one early pass of the survey area to set a proper motion baseline. %\todo{add numbers for depth and proposed coverage from whitepapers}
    \item[Solar system small bodies] Surveys or targeted observations with relatively short single exposures (less than 1 minute) to be coadded to a full depth of 24--27${}^{\mathrm{th}}$ magnitude, located within 2--10 degrees of the ecliptic plane to identify faint, small solar system bodies. Repeat observations of the same field on few-month timescales would yield orbits for these bodies.  
\end{description}

\begin{table}[]
\footnotesize
\centering
\renewcommand{\arraystretch}{1}

%would like to sort these in terms of themes probably?

    \begin{tabularx}{\textwidth}{C{0.3 in} L{3.5 in} L{1.4 in} L{1 in}}
    \Xhline{3\arrayrulewidth}
       \bf No. & \bf  Title &  \bf Contact Author & \bf Primary Theme \\
        \Xhline{2\arrayrulewidth}

11 & Roman Ultra Deep Field & Anton Koekemoer & Deep galaxies\\

\Xhline{\arrayrulewidth}

1 & Roman Multi-Tiered Surveys (RomanMTS) for Extragalactic Science	& Casey Papovich & Deep + grism \\
3 & Ultra Deep Field - Slitless Spectroscopy with Roman	& Sangeeta Malhotra & Deep + grism \\
7 & Establishing a Roman Ultradeep Legacy Field	& Brant Robertson & Deep + grism\\
9 & Roman Deep Survey &	Sean Bruton & Deep + grism \\
14 & Obscured AGN---Hiding High Growth at the Cosmic Noon	& Andreea Petric & Deep + grism\\

\Xhline{\arrayrulewidth}

5 & An Extended Time-Domain Survey (eTDS) to Detect High-z Transients, Trace the First Stars, and Probe the Epoch of Reionization & Ori Fox & Intermediate, wide-field \\ 
6 & KRONOS-E: Kp Roman Observations: New Opportunities from Surveying the Euclid deep-field	& Ranga-Ram Chary &  Intermediate wide-field + grism \\
16 & Roman Deep Survey of the Euclid Deep Fields	& Bahram Mobasher & Intermediate, wide-field \\
17 & Roman Cosmic Dawn Survey & Yuichi Harikane& Intermediate, wide-field\\
18 & Galaxy clusters in 3D: a deep imaging survey of the Virgo and Fornax clusters & Igor Chilingarian & Intermediate, wide-field \\
19 & Exploring the high redshift obscured accretion with Roman-WFIRST: A legacy survey leveraging on existing space and ground based facilities and tools.	& Nico Cappelluti & Intermediate, wide-field \\

\Xhline{\arrayrulewidth}

4 & The Roman Globular Cluster Time-Domain Survey: Probing the Oldest Stars and Exoplanets in the Milky Way	& Dan Huber & High cadence \\
8 & The TEMPO Survey: Probing Planet \& Satellite Formation with the Nancy Grace Roman Space Telescope	& Melinda Soares-Furtado & High cadence \\

\Xhline{\arrayrulewidth}

10 & The Fast and the Furious---a continuous cadence Galactic Plane survey with the Nancy Roman Space Telescope & Thomas Kupfer & Galactic plane \\
13 & A revolutionary new window into the dynamic and obscured Galactic plane with the Roman Space Telescope and Rubin Observatory & Kishalay De & Galactic plane\\
15 & Galactic Roman Infrared Plane Survey (GRIPS)	& Roberta Paladini & Galactic plane\\

\Xhline{\arrayrulewidth}

2 & Solar System Small Body Compositions using Roman WFI Spectrophotometry	 & Cristina Thomas & Solar System \\
12 & Roman Survey for Extreme TNOs (RoSET) & Bryan Holler & Solar System \\

\Xhline{\arrayrulewidth}
20 & Galactic Bulge Time Domain Survey & Andrew Rosenswie & No survey described\\
\Xhline{3\arrayrulewidth}
    \end{tabularx}
    \caption{White papers submitted to the Request for Information. Reference numbers were assigned in submission order; papers are grouped by theme (\S \ref{sec:themes}).}
    \label{tab:whitepapers}
\end{table}

\subsection{Deep fields to study the faintest objects} 
\label{ss:udf}
%text from summary slides
%Relevant science: EoR, early galaxy formation, AGN, Solar system

Deep exposures with space-based telescopes have a long history of extraordinary discoveries dating back to the original Hubble deep field. Today, astronomers studying galaxy formation in the early universe have coordinated to observe a handful of footprints, such as the CANDELS \citep{2011ApJS..197...35G, 2011ApJS..197...36K} and COSMOS \citep{2007ApJS..172....1S} fields, with deep observations spanning a wide range of wavelengths from radio to X-ray. This trove of multi-wavelength, coordinated data has become an important resource for the field of extragalactic astronomy; these data have also been mined for what they can tell us about foreground objects like distant Milky Way stars and solar system objects.

The long history of observing particular fields for extragalactic astronomy leads to a number of important arguments for early definition of a Roman ``ultra-deep field.'' Although the baseline of existing multiwavelength observations means that the set of fields likely to be observed for extragalactic science is already well defined, coordination is still needed since different fields have some fundamental advantages and disadvantages. In particular, early definition would provide a strong case in applications for much of the preparatory or follow-up spectroscopy that would be required. The committee noted that the time dedicated for Roman on Subaru, which is an important follow-up facility for this science case, will not be available by the start of the mission, but only several years into it. Early definition would also motivate groups to prepare for incorporating Roman data in multiwavelength analyses of deep fields; however, this type of development can already be funded through the Roman proposal call; the designation of an early definition survey does not come with its own funding a priori while the existing call does not require early definition. In fact, some familiarity with the performance of Roman's Wide-Field Imager may improve the results of deep observations that push the limits of the instrument. 

The committee thus perceived that while the gain in science \emph{capability} from early definition of this survey theme appeared modest, the potential \emph{utilization} of the data obtained in a Roman deep field (and thus the ultimate science outcomes) could still be large. Thus the gain from following a community-based approach to define such a survey, with the goal of maximizing this utilization, was perceived to be enormous. Having the GA proposal cycle drive field selection for such a survey will necessarily create insiders and outsiders even though data are all immediately public. For example, the choice of field location will determine which ground-based telescopes can perform follow-up observations (e.g., Northern vs. Southern Hemisphere, vs. equatorial fields accessible to both but with higher zodiacal background and more stringent Roman scheduling constraints). Observing time on many ground-based facilities is limited to the consortia that fund them, which then limits which groups can do some types of science. The committee perceived great importance in bringing as many people to the table as possible to define this type of survey, in order to ``average over'' political issues in favor of scientific value and create a more equitable dataset.

%For example, equatorial fields are accessible to ground based observatories in both hemispheres, and can be near the Ecliptic (most useful for the Solar System observations, as discussed in \S\S \ref{subsec:ss}). However, the zodiacal light is lower near the Ecliptic poles. If many rolls are important (as is needed for grism observations; see \S\S \ref{subsec:grism}) or if year-round accessibility is needed, then this motivates going to high Ecliptic latitudes. Time could also be split among 2--3 fields. 
%These and many other considerations together motivated our committee to conclude that a community based process for defining such a survey is imperative, since the survey's definition will determine the usefulness of this important public dataset. For example, the choice of field location will determine which ground-based telescopes can perform follow-up observations; observing time on many ground-based facilities is limited to the consortia that fund them, which then limits which groups can do some types of science.  Defining such a survey in a community-based process is an opportunity to create a more equitable dataset.

%On the other hand, it was not as compelling to the committee that such a survey be defined \textit{early}.  Thus the committee concluded that while the motivation for a community-based definition of an ultra-deep field is very strong, the motivation that this take place early is not as strong.

\subsection{Grism observations connecting feedback to galaxy formation} 
\label{ss:grism}
%text from summary slides
Several whitepapers presented compelling science cases for the study of early galaxy formation with Roman's grism, including star-forming galaxies at high redshift (now even more important thanks to new discoveries by the Webb telescope; \citealt{}) and the study of AGN at intermediate redshifts, which could resolve long-standing questions about the growth of supermassive black holes. The grism is a unique capability for Roman that complements existing ground- and space-based observatories, but fully leveraging its power requires complex deblending, stacking of spectra and different roll angles, and calibration (which requires both algorithm development {\em and} ancillary observations). The resulting deep spectroscopy provides information that \textit{cannot} be determined from imaging (even multiwavelength), and allows direct comparison between sources at high redshift and their lower-redshift, spectroscopically selected counterparts. 

Whitepapers proposing a survey using the grism presented several compelling justifications for early definition. The observations require specialized pipelines that differ materially from those developed for the core community surveys. Defining a grism survey early would provide motivation for this pipeline development and characterization of synthetic data. Some science cases using the grism also require specific ancillary observations at other wavelengths, for example to deblend the spectra from different sources; the algorithm to deblend by combining the new data sets still needs to be developed. The committee considered the need for early definition to motivate the calibration observations to be less compelling: in most cases the proposed targets are well-studied fields, for which some required ancillary data either exists already, while deep surveys over these fields currently being planned by Subaru/PFS and VLT/MOONS will be publicly available by the time they are needed.

The ultra-deep-field and grism whitepapers proposed surveys that spanned a continuum of combinations, devoting anywhere from zero to almost 100 percent of survey time to grism observations. The committee considered both the science cases and the combined motivations for early definition of a survey that included both imaging and grism observations to be more compelling than either case separately. 

%Proposals target well-studied fields, so much of the co-observing will likely happen anyway.

% A few additional thoughts from RCH:
% \begin{itemize}
% \item The grism provides a unique capability for Roman that complements existing ground- and space-based observatories. Fully leveraging this requires complex deblending, stacking of spectra and different roll angles, and calibration (which requires both algorithm development {\em and} ancillary observations. 
% \item The value of the deep spectroscopy is that it provides information that \textit{cannot} be determined from imaging (even multiwavelength) and it allows direct comparison between sources at high redshift and their lower-redshift, spectroscopically selected counterparts.
% \item The main question is -- what happens if this survey is {\em not} defined early? How long with the algorithm development and calibration take? And will the ancillary observations (in particular spectroscopy) happen anyway without the early definition?

% Some additional thoughts: There are some survey fields with sufficient depth and area being planned by Subaru/PFS and VLT/MOONS. The need for early definition to motivate these calibration observations is therefore less clear. 

% \end{itemize}

\subsection{Synergy at intermediate depths} %Intermediate depth, wide field of view
\label{ss:widefield}

Three whitepapers proposed observations over a relatively wide field of view (20--50 square degrees) at intermediate depths (26--29 magnitude) that leveraged synergy with other existing or planned surveys. These included a re-survey of the Roman Deep Fields in the K band (WP6), imaging over 7 bands in fields targeted by supernova searches (WP17), and imaging of the SDSS Stripe 82 fields surveyed by XMM (WP19). The target science for all three surveys was a better understanding of galaxy formation at very high redshifts within the epoch of reionization, a topic now made even more compelling by the recent discovery by the Webb telescope of relatively massive galaxies at high redshifts. The proposals made good use of the K-band filter planned for Roman, which is redder than anything offered by either Euclid or Rubin, and offered good justifications for their choices of field. 

The main justifications for defining such a survey early given by the whitepapers in this category, beyond the universal reasons cited in \S \ref{sec:rationale}, included coordination of observations by other surveys such as Rubin and Euclid, obtaining ground-based spectroscopy of the target galaxies, and the need to develop specialized analysis software. None of these were perceived by the committee to be compelling. In particular, the Euclid survey fields \citep{2022A&A...662A.112E} cited as one target for coordination have already been defined (an illustration of the importance of early definition, but in this case obviating the benefit defining the {\em Roman} field early). One whitepaper also pointed out the importance of early execution of one pass across the HLS field to obtain the longest possible baseline for proper motions of foreground stars, basically our only chance at studying the three-dimensional dynamics of the Milky Way's outskirts. The current reference survey provisionally defined for the HLS includes fields at the desired depths (although the reference survey pre-dated the addition of K band to the Roman mission). Thus the committee viewed decisions about the location and filters used for these observations to be primarily the responsibility of the CCS definition committee for the HLS, rather than indication of a need for defining another survey at these depths. We therefore recommend that the considerations identified in these whitepapers be passed on to the CCS definition committee when it is formed, and taken into account during the definition of the HLS.

\subsection{The variable sky} %High time-cadence, small field of view
\label{ss:timedomain}
%text from summary slides
The subset of whitepapers dealing with high time-cadence, small field of view observations (i.e., \textit{not} a large footprint of the Galactic plane) outlined compelling science cases spanning stellar astrophysics. A time-domain survey of globular clusters could detect planets around old stars and study stellar structure at low metallicity via asteroseismology, exploiting Roman's wide field to survey across the entire HR diagram and simultaneously map mass segregation across an entire cluster. A time-domain survey of a young star-forming region could conversely study star formation in situ and planets around young stars.  Proposals even extended to high-redshift searches for supernovae from Population III stars.

The justifications for early definition put forward in whitepapers for this type of survey articulated a very strong need for early \textit{execution}, to obtain baselines for proper motion measurements or follow up potential Population III supernovae. There is also a need, relatively unique to this type of survey, to optimize the scheduling of simultaneous or timed observations. The committee recognized the potential of this work for interesting technical achievements but felt that the science applications lacked the comprehensive breadth of some other themes. In addition, several whitepapers pointed out the need to develop and test software to deal with saturation around very bright stars and to analyze crowded fields. These justifications for early definition were less compelling to the committee, since pixel-fitting for saturation will be developed for the GBTDS and a crowded-field photometry pipeline is in development by one of the recently funded Roman Large WFS proposal teams. The committee considered that the motivation for early definition was therefore driven by the need for early execution, and suggests that such a survey could be considered for Cycle 1.

\subsection{The Galactic plane}
\label{ss:galacticplane}

Two white papers described the broad range of science that could be explored with a Roman survey of the Galactic plane. While Gaia has revolutionized our understanding of the outer Milky Way and solar neighborhood, mapping the inner Galaxy requires a combination of capabilities that is unique to Roman: a wide field of view, infrared wavelength coverage to penetrate dust, and high spatial resolution to deal with crowding. The committee recognized that such a survey would truly be the first of its kind; the only comparable views were those offered by the GLIMPSE surveys carried out on Spitzer's IRAC instrument, which had a depth of 13-15.5 mag at 8.0-3.6 microns respectively \citep{Churchwell_2009} and a resolution of 2 arcsec. In comparison, the proposed surveys would reach depths of 23--25.5 mag at 1.06--2.13 microns with Roman's resolution of 0.1 arcsec, an improvement of a factor 10 in depth and 20 in resolution. Roman's wide field of view makes it the only observatory that could achieve this depth in a full survey of the inner Galaxy. Besides complementing existing and planned surveys, these observations would also provide an important link between Galactic and extragalactic studies carried out with Roman. The committee also appreciated the complementarity between the core community surveys, which are driven by science pertaining to exoplanets and cosmology, and a survey that would feature Galactic astronomy---a study at intermediate length scales to either of the core surveys, and spanning many of the areas of interest not covered by their science cases.

\begin{figure}[!ht]
    \centering
    \includegraphics[width=\textwidth]{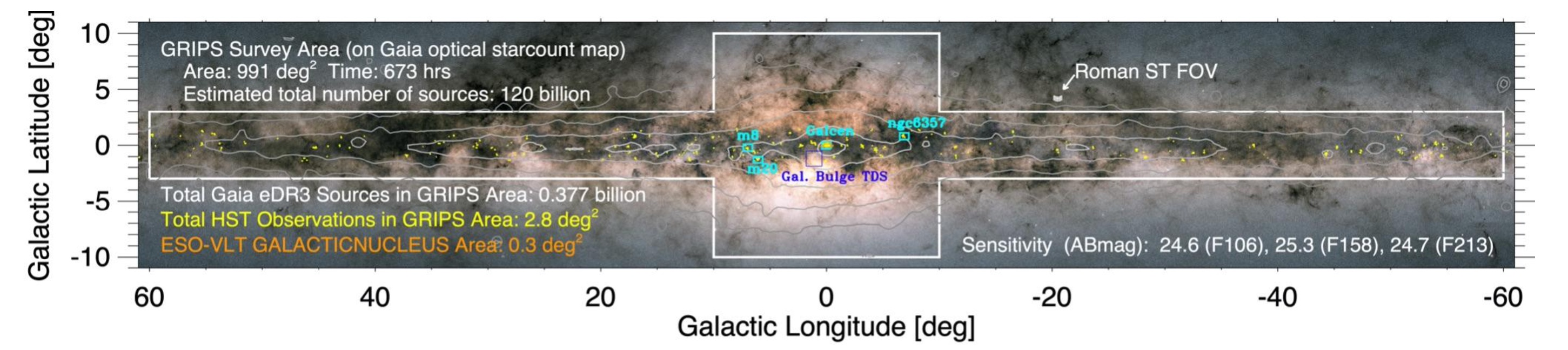}
    \caption{Proposed observational plan (white solid lines) for a galactic plane survey from WP15, overlaid on an optical Gaia starcount map. Grey contours show COBE/DIRBE 4.9 $\mu$m emission at 1,2,4,8, and 16 MJy/sr, outlining the stellar disk and bulge. High angular resolution coverage by HST (yellow) and the ESO-VLT GALACTICNUCLEUS program (red) are also shown, as is the field for the Roman Galactic Bulge Time Domain Survey (blue). Star-forming fields in the Galactic Center, NGC 6357, M8, and M20 are noted in cyan.}
    \label{fig:planesurvey}
\end{figure}

The committee considered an infrared survey of the inner Galaxy with these parameters to be unprecedented and potentially transformational, with impact across a very broad range of disciplines. The committee considered the potential inherent in this largely unexplored discovery space to be very large. Scientific applications would include studies of the Galaxy's structure and dynamics in stars and dust, the environmental dependence of star formation, the coeval evolution of the Galactic nucleus and its resident supermassive black hole, the evolution and properties of flaring and variable stars, compact-object binaries, and the potential for detecting Galactic supernovae. Nearly all of this science requires a time-domain component, to measure proper motions and other photometric and astrometric variability. This type of survey also has strong synergies with Rubin, which could provide coverage at visible wavelengths and high cadence complementary to Roman's capabilities; combined catalogs would deliver deep multi-band star and dust maps, proper motions and parallaxes, dense light curves, astrometric trajectory curves, and excellent understanding of completeness even in crowded regions. 

The committee considered the justifications for early definition, as laid out by the whitepapers proposing a Galactic plane survey, to be compelling. Such a survey would require a high level of coordination between stakeholders across multiple disparate subfields of astrophysics that have traditionally interacted relatively little, including not only survey definition but also the development of coordinated surveys at other wavelengths (radio, optical and X-ray) to amplify science yields. The early survey definition process would ensure that survey area, cadence, and outcomes are defined collectively by this broad community and that the full breadth of science can be realized. Evaluating tradeoffs between cadence and plane coverage will require a larger than usual time investment given the variety of relevant science cases. Analysis pipelines that can cross-match targets in crowded fields with mixed spatial resolutions between Roman and ancillary observations would differ substantially from the pipelines needed for the existing core community surveys, and thus require longer lead times. Many of the science cases require proper motion information, for which the precision is maximized if the plane survey is started early in the Roman lifecycle, which the committee viewed as a compelling argument for early execution of at least part of such a survey.
Finally, defining a Roman Galactic plane survey early may also influence the Rubin survey strategy, which will be finalized over the next 1--2 years. 

The committee also noted the potential for public engagement from the visualizations of our home Galaxy that could be created from such a survey, in the spirit of the Gaia and PHAT surveys: not only stunning imaging that could place already-famous icons like HST's ``pillars of creation'' into their full context via extreme zooms, but also movies showing the motions of stars and gas as they interact with each other. Although the committee did not consider public engagement a good justification for early definition on its own, they felt that combined with other motivations it was a notable advantage for this case.

\subsection{The Solar System}
\label{ss:solarsystem}
Roman will be able to discover faint bodies in our Solar System that are more distant, rarer and smaller than ever before. Whitepapers proposing a discovery survey for the Solar System pointed out that these new small bodies will carry new and different information about the formation of the Solar System, beyond what we have ever been able to access. The opening of this new discovery space would be transformational for Solar System studies; for example potentially resolving questions about whether small bodies have a characteristic size, discriminating between several leading theories of Solar System formation, and searching for evidence of distant planets. Roman's wide field is uniquely capable of making these discoveries; it can not only find these faint bodies but obtain their orbits through repeated observations.

Whitepapers articulated several justifications for defining a Solar System survey early. The most compelling, in the committee's view, was the importance of field selection for such a pathfinding survey, which would open up a new discovery space that could subsequently spawn many GA proposals. Given the breadth of applications within the Solar System community, the committee recognized this as another theme where carrying out a community-based survey definition process could be an important opportunity to create a more equitable outcome. 

Less compelling as motivations for early definition were development of studies to optimize Roman's capabilities as a small-body discovery machine. These included working out how to mitigate cosmic rays for long co-added exposures and deal with moving objects, and modeling to optimize the survey. The committee noted that substantial work on identifying moving objects in deep exposures has already taken place \citep[e.g.][]{Bernstein.2004}, while development of software pipelines can be funded through the Roman Wide-field Science proposal calls without the need to first define a survey. The committee also noted that the requirements for Solar System observations often dovetailed with those for ultra-deep fields (\S\S \ref{ss:udf}), and that at least one of the extragalactic fields with multiwavelength observations is at relatively low ecliptic latitude (the COSMOS field). It may well be worth investigating these synergies for future Roman observations.

%text from summary slides
%Faint bodies will be discovered with Roman = more distant, rarer \& smaller. These encode new/different info about SS formation, e.g. was there a characteristic size (i.e. break in power law)? 
 %-> repeated observations w/Roman = orbits. Wide field is key.

%justification for early def:
%Cosmic ray mitigation for long co-adds (see RoSEt proposal)
%Dealing w/moving objects (mostly done already)
%Modeling for optimization (see Thomas)

%Potentially, optimizing co-observations with extragalactic programs? (see notes)

%None appeared extremely compelling (though adjust for how long Matt thinks things take!) 

%Individual exposure times tend to be much shorter (less than 1 minute) for surveys targeting main belt objects to avoid streaking, but longer for TNOS (and coadd plans in all cases)

\section{Findings and Recommendations}

\subsection{Recommendations for an early definition survey}

{\bf The committee found that there was sufficient justification to execute an early-definition survey for the Roman telescope.} Several of the categories of proposed surveys described in the submitted whitepapers made a convincing case for early definition. Even the survey themes that the committee perceived to have a less convincing case for early definition both illustrated the power of the Roman telescope for discoveries across a very broad range of science cases and pointed out important considerations that should be taken into account in defining the core community surveys.  Below we summarize our ranking and include some notes on each theme.

\begin{tabularx}{\textwidth}{C{0.4 in} L{2.1 in} L{4.4in}}
    \Xhline{3\arrayrulewidth}
       \bf Rank & \bf  Theme &  \bf Notes \\
        \Xhline{2\arrayrulewidth}
1 & Galactic plane  & See \S\S \ref{ss:galacticplane} for a discussion of suggested parameters. \\
2 & Deep field + grism & Strongly recommend community process even if not early. \\
\Xhline{\arrayrulewidth}
3 & Time domain survey & Compelling motivation for early execution; should strongly consider for Cycle 1. Technical needs should be communicated with the core community survey definition teams. \\
3 & Solar system & New discovery space, but less time-pressure than other areas. \\
\Xhline{\arrayrulewidth}
NR & Wide field, intermediate depth & Impacts of choices on science should be communicated to the high-latitude core community survey definition team.\\
    \Xhline{2\arrayrulewidth}
\end{tabularx}

Given this ranking the committee also recommended several aspects of the scope for a Galactic Plane survey for consideration by the committee defining this survey. The definition committee should consider which areas of high extinction and crowding to include (especially regions where Gaia is most challenged), whether or not to include the Galactic bulge, and what aspects of the time domain are most compelling. 

\vspace{3mm}

\subsection{Other findings}

Below we summarize specific findings and recommendations ancillary to our main charge.
\begin{itemize}
    \item The definition of an ultra deep survey should be done via a community process. This approach should also be considered for the other themes studied here, since all have broad applicability across many compelling science cases.
    \item There is at least a possibility of synergy between a deep extragalactic field and some Solar System science goals. Some well studied deep extragalactic fields are at relatively low ecliptic latitude and so may be effective for Solar System science. Zodiacal foreground may be an issue depending on the precise science goal of the deep observations, but this can be mitigated somewhat by timing of the observations. This is worth investigating further by a future committee.
    \item Many compelling cases were identified in time-domain (WP4, WP5, WP8) and galactic plane (WP10, WP13, WP15) whitepapers for early execution or scheduling of HLS fields, for e.g. Population III SNe follow-up and establishing a proper motion baseline. This should be communicated to the team defining this CCS.
    \item Technical needs articulated in these whitepapers, such as cosmic ray and saturation mitigation and grism and crowded-field photometry pipelines, often recurred across themes. These should be communicated to the Project Infrastructure Teams and science teams tasked with developing the technical foundations for the core community surveys, for better clarity on how to accommodate these science-driven needs within the requirements for the core surveys.
        
\end{itemize}

One of the most important themes that emerged as the committee deliberated was the power of a community-based process to define key surveys to be carried out by Roman in a more equitable fashion than competed proposals from individual teams. Though Roman's data will be immediately public, the ability to make full scientific use of this data is not guaranteed to be equitable, so it is worth considering how to lower barriers to access. Of course not all Roman observations would benefit from this approach, but for surveys with the broad reach described in the submitted whitepapers, and for which the survey definition is likely to create clear advantages or disadvantages for scientists based on things like other telescope access for follow-up observations, the committee considered a community approach to be a potentially powerful tool for increasing equity and maximizing the potential for scientific discovery.

\bibliographystyle{aasjournal}
\bibliography{refs}

\end{document}